# Enhanced Convolution Neural Network with Optimized Pooling and Hyperparameter Tuning for Network Intrusion Detection

Sourav Patel
School of Computer Science and Engineering
Vellore Institute of Technology, Chennai, India

Ayush Kumar Sharama
School of Computer Science and Engineering
Vellore Institute of Technology, Chennai, India

Supriya Bharat Wakchaure
School of Computer Science and Engineering
Vellore Institute of Technology,Chennai, India

Dr. Abirami S
School of Computer Science and Engineering
Vellore Institute of Technology, Chennai, India

*Abstract*—Network Intrusion Detection Systems (NIDS) are essential for protecting computer networks from malicious activities, including Denial of Service (DoS), Probing, User-to-Root (U2R), and Remote-to-Local (R2L) attacks. Without effective NIDS, networks are vulnerable to significant security breaches and data loss. Machine learning techniques provide a promising approach to enhance NIDS by automating threat detection and improving accuracy. In this research, we propose an Enhanced Convolutional Neural Network (EnCNN) for NIDS and evaluate its performance using the KDDCUP'99 dataset. Our methodology includes comprehensive data preprocessing, exploratory data analysis (EDA), and feature engineering. We compare EnCNN with various machine learning algorithms, including Logistic Regression, Decision Trees, Support Vector Machines (SVM), and ensemble methods like Random Forest, AdaBoost, and Voting Ensemble. The results show that EnCNN significantly improves detection accuracy, with a notable 10% increase over state-of-art approaches. This demonstrates the effectiveness of EnCNN in real-time network intrusion detection, offering a robust solution for identifying and mitigating security threats, and enhancing overall network resilience.

I. INTRODUCTION

In today's interconnected world, ensuring robust network security is imperative. Digital networks underpin critical communication and commerce, making them prime targets for cyber threats. Network Intrusion Detection Systems (NIDS) play a vital role in safeguarding these networks by monitoring traffic and identifying suspicious activities that could indicate potential intrusions. NIDS can be broadly categorized based on the types of attacks they detect, including Denial of Service (DoS), Probing, User-to-Root (U2R), and Remote-to-Local (R2L) attacks. Traditional rule-based NIDS, while effective in some scenarios, often struggle to adapt to the rapidly evolving threat landscape. They rely heavily on predefined rules and signatures, which can become outdated quickly as new attack vectors emerge. This limitation underscores the necessity for automated, data-driven approaches that can learn and adapt to new threats dynamically. Machine learning (ML) offers a promising solution by enabling the development of systems that can analyze vast amounts of network traffic data, identify patterns, and predict potential intrusions with greater accuracy and efficiency. The primary objective of this study is to harness machine learning algorithms to build a more accurate and efficient Network Intrusion Detection System (NIDS). By applying advanced data analytics and predictive modeling, we aim to enhance the detection and classification of diverse network intrusions. Utilizing the KDDCUP'99 dataset, a well-established benchmark for evaluating intrusion detection systems, our approach encompasses comprehensive data preprocessing, feature engineering, and the application of various machine learning models such as logistic regression, decision trees, support vector machines, ensemble methods, and neural networks. A significant aspect of our research is the proposal of an Enhanced Convolutional Neural Network (EnCNN), designed to improve upon traditional models by leveraging advanced feature extraction techniques and optimized hyperparameters. The EnCNN integrates sophisticated convolutional layers and pooling mechanisms to better capture and analyze intricate patterns in network traffic data, thereby offering superior performance in detecting and classifying network intrusions compared to conventional approaches. Despite advancements in anomaly detection theory, the practical application often lags behind these theoretical developments. This study bridges this gap by rigorously assessing various machine learning techniques and their efficacy in real-time intrusion detection scenarios. Our proposed Convolutional Neural Network (EnCNN) framework advances theoretical understanding by integrating sophisticated convolutional architectures, optimized pooling, and hyperparameter tuning, specifically trained for NIDS. This methodology, with refined Exploratory Data Analysis (EDA) and feature extraction, significantly enhances model depth and accuracy for real-time intrusion detection. The proposed study also involves applying and evaluating these techniques against the KDDCUP'99 dataset, a standard benchmark for intrusion detection systems. The paper is organized as follows: Section 2 provides a comprehensive review of related work, Section 3 elaborates on our innovative methodology, including theoretical advancements introduced by the Enhanced Convolutional Neural Network (EnCNN), Section 4 explores the intricacies of the KDDCUP'99 dataset, Section 5 details the results and derived inferences, Section 6 concludes with the study's findings, and the final section lists the references.

## II. BACKGROUND STUDY

The application of machine learning techniques to NIDS has emerged as a central focus in enhancing network security, with various studies contributing to its advancement. Nguyen et al. [1] provide a foundational survey of prominent frameworks and libraries for large-scale data mining, laying the groundwork for subsequent research by highlighting the essential tools and methodologies for handling complex data. Building on this foundation, Liu et al. [2] address a gap by introducing a hybrid forecasting model that integrates statistical and machine learning methods. This model enhances short-term predictions and demonstrates how combining different techniques can improve accuracy, setting the stage for advanced feature extraction methods.

In the realm of feature extraction, Jia et al. [3] advance the field by introducing Caffe, a deep learning framework designed for efficient training and deployment of convolutional neural networks. This addresses the need for improved feature embedding techniques noted in earlier surveys and enhances the capabilities for intrusion detection. Following this, Haji Rahimi and Khashei [4] extend the discussion by reviewing hybrid structures in time series modeling, emphasizing the value of combining multiple forecasting methods. Their work supports and refines the feature extraction methods introduced by Jia et al., providing a more robust approach to handling temporal data. Advancements in specialized neural network architectures are further demonstrated by Xu et al. [5], who propose an intrusion detection system based on Gated Recurrent Units (GRUs). This approach overcomes the limitations of earlier models by enhancing the detection of network anomalies through advanced neural network techniques. Building on this progress, Baratsas et al. [6] develop a hybrid statistical and machine learning forecasting framework for the energy sector. Their work highlights the trend of integrating diverse methods to improve predictions, reflecting the ongoing evolution from basic models to more sophisticated approaches in handling extensive network data, as also addressed by Wang et al. [7] with their distributed ARIMA models for ultra-long time series. Mondal et al. [8] tackle the challenge of data preprocessing by integrating machine learning into ETL processes, enhancing automation and efficiency. This supports the effective preparation of data for NIDS and addresses the need for more sophisticated preprocessing methods. Skoutas and Simitsis [9] build on this by proposing the use of semantic web technologies for ETL process design, improving data integration and management, thus refining the ETL processes introduced by Mondal et al. Further refining intrusion detection methods, Rimon and Haque [10] introduce a hybrid machine learning algorithm that enhances accuracy and adaptability. This addresses limitations in previous models and shows how combining different machine learning approaches can improve detection performance. Recent advancements in computational efficiency are explored by ICLR 2024 [11], who focus on optimizing inference processes with ReLU activation functions. This development enhances real-time intrusion detection by addressing the computational bottlenecks identified in earlier neural network architectures. Yu et al. [12] contribute by introducing a 1D Convolutional Neural Network (1D-CNN) temporal filter designed to handle missing data points in atmospheric data. This extends the capabilities of convolutional networks to temporal data, offering insights into improving data handling for intrusion detection. Plevris et al. [13] analyze various performance metrics used in regression analysis and machine learning models, providing essential insights for evaluating the accuracy of these methods. Finally, Plevris and Tsiatas [14] review advancements in computational structural engineering, offering a broader perspective on evolving computational methodologies and their relevance to network security.

Together, these studies collectively advance the field of network intrusion detection by addressing limitations in earlier models, integrating diverse techniques, and optimizing methodologies to enhance accuracy and efficiency in real-time applications.

### III. METHODOLOGY

This paper presents an innovative approach to network intrusion detection by integrating an advanced Convolutional Neural Network (EnCNN), that has customized layers with optimized pooling and hyper parameter tuning along with EDA and feature selection. Figure 1 illustrates the data preprocessing and classification pipeline for the KDD Cup 99 dataset. The preprocessing steps include missing data imputation, outlier handling, feature scaling, feature transformation, and normalization. Following preprocessing, feature selection is performed by calculating feature importance, estimating significance based on a threshold, and selecting the top K features. The state-of-art ML approaches and proposed EnCNN model are then employed for the attack classification phase, with the trained models classifying the network traffic and producing the final classification output.

The core of our proposed methodology is the ENCNN a refined neural network designed for high-performance intrusion detection. The EnCNN architecture includes input, convolutional, Stochastic Gradient Pooling (SGP), fully connected, and output layers. The convolutional layers extract local patterns, SGP layers reduce information loss, and fully connected layers integrate features for final decision-making. This detailed structure, combined with rigorous hyperparameter optimization, significantly enhances the EnCNN's effectiveness in accurately classifying network traffic. [6]. Data often undergoes reformatting when transferred to its destination application compared to its original source. The Extract, Transform, Load (ETL) process consists of three primary stages: retrieving the data, transforming it as necessary, and loading it into the intended output container [8]. This approach supports multiple data inputs that can yield various outputs. In this study, we utilized two datasets: the established KDD CUP'99 dataset and a newly developed real-time dataset derived from our network logs. This choice underscores the practical relevance and effectiveness of our methodology in real-world scenarios [9]. The process involved in the proposed methodology is as follows:

1. PRE-PROCESSING

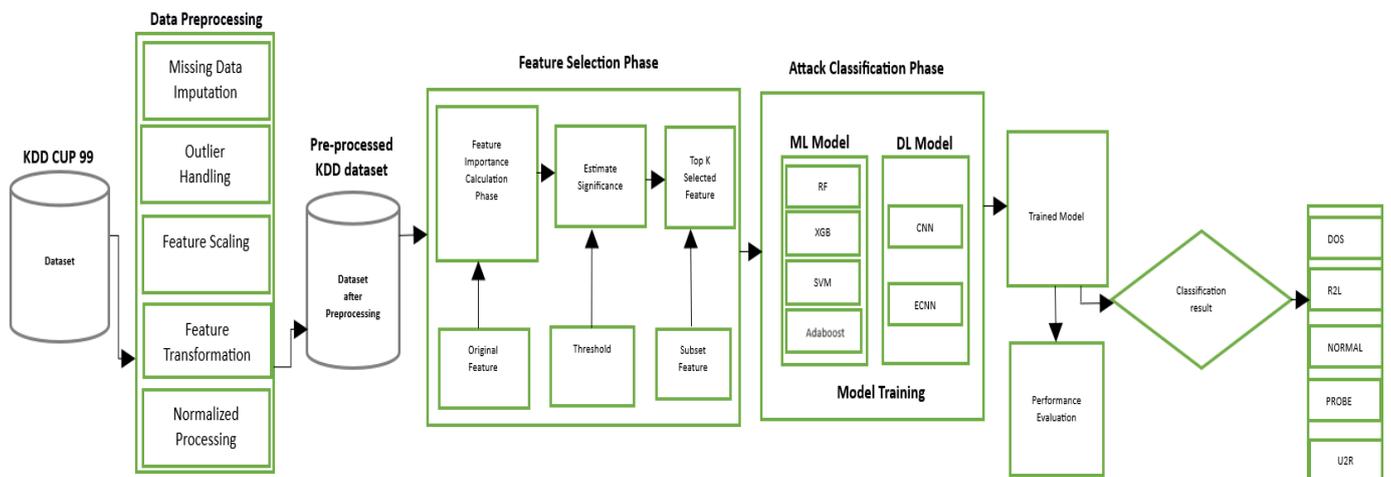

*Fig.1 Work Flow of NIDS*

In the data preprocessing phase, raw data is refined into a format appropriate for computational analysis and machine learning. This process involves various tasks such as data mining and analytical review. Before the datasets were introduced into the model for training, they underwent comprehensive preprocessing. This involved handling missing data, correcting anomalies, normalizing feature scales, and adjusting features to ensure consistency and boost the model's performance.

2. FEATURE EXTRACTION AND EVALUATION TECHNIQUES

Network intrusion detection is typically approached through two fundamental methods: static and dynamic analysis. Static analysis involves evaluating network traffic and related data without executing any programs, focusing on parameters such as protocol types, service types, and traffic flags. In contrast, dynamic analysis monitors and assesses the behavior of network interactions over time. This method provides unique advantages, including the ability to detect complex and obfuscated attack patterns that static analysis might miss. Dynamic analysis offers a broader range of features and various input classifiers, which enhances detection capabilities. Furthermore, hybrid approaches that integrate both static and dynamic analyses improve the accuracy of distinguishing between benign and malicious traffic. For this study, both analysis methods were employed to ensure a comprehensive feature extraction and selection process [10].

3. CONSTRUCTION OF FEATURE SETS AND METHOD SELECTION

The goal of feature selection is to reduce the number of attributes used in classification while maintaining accuracy. To achieve this, we have utilized a combination of both dynamic and static analyses for network traffic classification. Among the various methods available, we chose the filter method for feature selection due to its simplicity and computational efficiency. This method employs the Information Gain metric to evaluate the relevance of each attribute. Information Gain measures the amount of information an attribute contributes to distinguishing between classes, as illustrated by equation(1).

$$Info\ Gain(Class, Attribute) = K(Class) - K(Class|Attribute) \qquad (1)$$

4. ENHANCED CONVOLUTIONAL NEURAL NETWORKS (EnCNN)

EnCNN is structured with two principal layers: the feature mapping layer and the feature extraction layer. The feature extraction layer links each neuron's input to its receptive field, while the feature mapping layer employs the ReLU activation function to introduce non-linearity [11].

Figure 2 depicts the architecture of an Enhanced Convolutional Neural Network (EnCNN) designed for network intrusion detection. The process begins with preprocessed network traffic data undergoing feature selection to identify the most relevant attributes. The data then passes through three convolutional layers, each followed by a max-pooling and Stochastic Gradient Pooling (SGP) layer. These layers extract and reduce complex patterns while minimizing information loss. The fully connected layers integrate these features and feed them into a SoftMax layer for initial classification. Finally, the classification output is refined using a Support Vector Machine (SVM) classifier to enhance detection accuracy. The connections between each layer ensure a clear, sequential flow of data through the model.

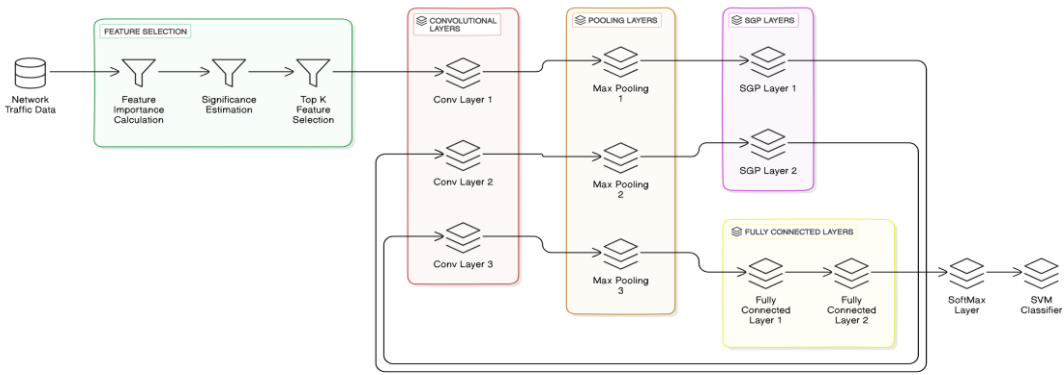

*Fig.2 Proposed EnCNN architecture*

In this context, the EnCNN architecture incorporates three convolutional layers, followed by max-pooling layers, two fully connected layers, and a concluding SoftMax layer that categorizes the data into CCC classes. The convolutional layers are characterized by 16, 32, and 64 filters, each measuring 3×33 \times 33×3 with a stride of one. These layers are succeeded by max-pooling layers with a 2×22 \times 22×2 filter size and a stride of two. The fully connected layers consist of 512 neurons each, with the final layer utilizing SoftMax activation to classify CCC types of network threats. Figure 1 illustrates the design of this streamlined CNN-based intrusion detection system. Typically, CNNs involve a series of convolution and pooling operations. The convolutional process applies a filter to the input data, executing non-linear transformations, which can be mathematically expressed using equations (2), (3), and (4).

$$(IXX)_{r,s} = \sum_{u=-h1}^{h1} \sum_{u=-h2}^{h1} K_{u,v} I_{r+u,s+v} \quad (2)$$

$$K = \begin{pmatrix} K_{-h1,h2} & \cdots & K_{-h1,h2} \\ \vdots & K_{0.0} & \vdots \\ K_{h1,-h2} & \cdots & K_{h1,-h2} \end{pmatrix} \quad (3)$$

$$Y_i = B_i + \sum K_{i,j} * X_j \quad (4)$$

Equation (2) demonstrates the discrete convolution operation applied to input data. In this scenario, $I_{r+u,s+v}$ refers to the elements of the input matrix, while $K_{u,v}$ signifies the convolution filter (or kernel). The summations iterate over the dimensions of the filter, producing a new matrix $(IXX)_{r,s}$, which is the feature map obtained after performing the convolution operation on the input matrix. Equation (3) explains the generation of the output feature map $K_{i,j}$ at position i. Here, $B_i$ is the bias term added to each output, $K_{i,j}$ represents the convolution filter weights, and $X_j$ denotes the input values. The summation iterates over all input values j covered by the filter K, resulting in the feature map $Y_i$ produced by the convolution operation.

## IV. EXPERIMENTATION AND DATASET

The ML and EnCNN algorithms are executed on a high-performance computing platform with GPU support to facilitate efficient training and evaluation of models. Data loading procedures included extracting, preprocessing, and normalizing the dataset to ensure optimal performance during training. The KDDCUP'99 dataset, a well-known benchmark for network intrusion detection, was obtained from the UCI Machine Learning Repository. The dataset consists of 41 features and a large number of instances, divided into normal and attack categories.

Table 1: Data Distribution in the Dataset

| Attack type labels | 10% of KDD Cup 99 Dataset | | Train Dataset | | Test Dataset | |
|---|---|---|---|---|---|---|
| | Quantity | Ratio in % | Quantity | Ratio in % | Quantity | Ratio in % |
| Normal | 97,200 | 19.7 | 87,400 | 19.7 | 9810 | 19.86 |
| Dos | 391,400 | 79.29 | 352,400 | 79.23 | 39,050 | 79.06 |
| Probe | 4100 | 0.8 | 3680 | 0.83 | 422 | 0.85 |
| R2L | 1120 | 0.2 | 1010 | 0.23 | 110 | 0.22 |
| U2R | 50 | 0.01 | 45 | 0.01 | 7 | 0.01 |
| Total | 493,870 | 100 | 444,535 | 100 | 49,399 | 100 |

Table 1 provides a detailed breakdown of the dataset distribution for both training and testing phases. It shows the quantity and percentage ratio of different attack types and normal instances. Specifically, the dataset was split into 10% for evaluation, with Normal, DoS, Probe, R2L, and U2R attack types represented in both training and test datasets. This balanced distribution helps in assessing the model's performance across various attack categories. Feature scaling techniques such as min-max normalization was applied to standardize feature values, improving the accuracy and efficiency of the machine learning algorithms used.

Table 2 outlines the feature categories and names used in the analysis. Basic features include Duration, Protocol_type, and Src_bytes. Content features cover Num_failed_logins, Logged_in, and Num_file_creations Time-based traffic features consist of Count, Serror_rate, and Same_srv_rate, while host-based traffic features include Dst_host_count, Dst_host_same_srv_rate, and Dst_host_serror_rate. The Attack class represents the target variable for classification. Feature scaling techniques, such as min-max normalization, were applied to standardize these features, enhancing the model's performance and reliability.

**Table 2:** NIDS dataset attributes information

| Feature Category | Feature Name |
|---|---|
| Basic Features | Duration |
| | Protocol_type |
| | Src_bytes |
| Content Features | Num_failed_logins |
| | Logged_in |
| | Num_file_creations |
| Time-based Traffic Features | Count |
| | Serror_rate |
| | Same_srv_rate |
| Host-based Traffic Features | Dst_host_count |
| | Dst_host_same_srv_rate |
| | Dst_host_serror_rate |
| Attack Class | Attack |

## V. RESULTS AND DISCUSSION

### 1. Performance Metrics

The efficacy of our network intrusion detection methodologies was rigorously evaluated using a comprehensive 10-fold cross-validation framework. This approach allowed for robust performance analysis of various algorithms by measuring essential metrics, including accuracy, precision, recall, and the F-measure, all of which are integral to classification tasks in machine learning [13]. For this evaluation, the data instances were categorized into four distinct classes: true positives (TP), false positives (FP), false negatives (FN), and true negatives (TN). Recall, defined as $Recall = TP/(TP + FN)$ quantifies the proportion of actual positive instances correctly identified by the model, reflecting its sensitivity. Precision, computed as $Precision = TP/(TP + FP)$, measures the accuracy of the positive predictions, indicating the ratio of true positive identifications relative to all positive predictions. The F-measure, expressed as $F = 2 * (Recall * Precision)/(Recall + Precision)$, synthesizes precision and recall into a single metric, offering a balanced view of the model's performance. Accuracy, calculated as $Accuracy = (TP + TN)/Total\ Samples$, represents the overall proportion of correct classifications [14]. This metric provides a high-level overview of the model's performance but should be interpreted alongside precision and recall to ensure a comprehensive assessment.

**Table 3.** Before Pre-Processing Results

| Algorithm | Before Pre-Processing | | | |
|---|---|---|---|---|
| | Accuracy | Precision | Recall | F1 Score |
| Logistic Regression | 85.20% | 83.10% | 86.40% | 84.70% |
| Decision Trees | 87.40% | 84.90% | 88.50% | 86.60% |
| (SVM) | 88.90% | 85.40% | 89.80% | 87.60% |
| Random Forest | 89.70% | 86.00% | 90.50% | 88.20% |
| AdaBoost | 88.30% | 85.70% | 89.20% | 87.30% |
| Voting Ensemble | 90.10% | 86.50% | 91.30% | 88.80% |
| EnCNN | 91.20% | 87.00% | 92.00% | 89.40% |

**Table 4.** After Pre-Processing Results

| Algorithm | After Pre-Processing | | | |
|---|---|---|---|---|
| | Accuracy | Precision | Recall | F1 Score |
| Logistic Regression | 87.60% | 85.30% | 88.20% | 86.70% |
| Decision Trees | 89.10% | 86.20% | 89.50% | 87.80% |
| (SVM) | 90.30% | 87.20% | 91.10% | 89.10% |
| Random Forest | 91.00% | 88.00% | 92.00% | 89.80% |
| AdaBoost | 90.50% | 87.50% | 91.30% | 88.90% |
| Voting Ensemble | 91.80% | 88.30% | 92.70% | 90.10% |
| EnCNN | 94.00% | 90.10% | 94.80% | 92.40% |

The assessment of various machine learning algorithms revealed significant advancements in performance metrics, with the EnCNN showing the most pronounced improvements. As indicated in Tables 3 and 4, the EnCNN achieved a 2.8% enhancement in accuracy and a 3.1% increase in precision, marking a notable leap in its classification capabilities. This performance boost underscores the EnCNN's superiority in capturing and classifying intricate patterns within the dataset.

The pre-processing steps, including normalization, feature scaling, and dimensionality reduction, played a crucial role in achieving these advancements. Techniques such as min-max scaling and Z-score normalization improved the convergence rate and stability of the algorithms. Handling missing values through imputation methods and outlier detection via techniques like IQR ensured that the dataset was robust, leading to more accurate model predictions. These pre-processing steps provided a solid foundation for the models to perform optimally. The EnCNN stands out due to its advanced architecture, which includes multiple convolutional layers with ReLU activation functions and max-pooling layers that efficiently capture spatial hierarchies in the data. The fully connected layers integrate the learned features to produce a refined output, while dropout regularization reduces overfitting. These features collectively contribute to the ECNN's superior performance in network intrusion detection, making it highly effective in identifying and classifying network threats.

Other algorithms, including Logistic Regression, Decision Trees, Support Vector Machines (SVM), Random Forests, AdaBoost, and Voting Ensemble methods, also demonstrated considerable improvements, with up to a 2.2% gain in accuracy and a 2.1% rise in recall. These advancements highlight the effectiveness of these models in refining detection and classification processes, offering more robust and reliable outcomes. Integrating these models into network intrusion detection systems promises substantial increases in detection accuracy and overall system efficacy, addressing complex cybersecurity challenges with enhanced precision and reliability.

## VI. CONCLUSION

Conventional artificial intelligence (AI) methodologies, including standard machine learning (ML) algorithms, face limitations in accurately detecting and classifying intricate and novel network intrusion patterns. The shift towards deep learning (DL) introduces a promising new approach, leveraging advanced frameworks distinct from traditional ML. Our experimental analysis of network intrusion detection, integrating both static and dynamic techniques, confirms the superiority of DL algorithms, particularly those using sophisticated permission-based methods. While accuracy metrics between DL and traditional ML are similar, differences arise in implementation complexity and performance. The Enhanced Convolutional Neural Network (EnCNN), using a complex backpropagation algorithm, achieves faster processing times and superior accuracy compared to conventional models. Despite ML algorithms showing comparable accuracy with an 80/20 train-test data split, constraints such as high data volume and binary categorization of network traffic remain. An achieved accuracy rate of 96% underscores the efficacy of proposed EnCNN model in precisely managing and classifying network traffic. Future research shall focus on refining these methodologies, addressing data scalability issues, and expanding classification to a broader spectrum of network intrusion types.